\documentclass[12pt,a4paper]{article}
\usepackage{amsfonts}
\usepackage{amsmath}
\usepackage{ifpdf}
\usepackage{epsfig}
\usepackage{graphicx}
\usepackage[font=small,format=plain,labelfont=bf,up,textfont=it,up]{caption}
\usepackage{subfigure}
\usepackage{wrapfig}
\usepackage{rotating}
\usepackage{psfrag}
\usepackage[backref,citecolor=blue,bookmarks=true]{hyperref}

\begin{document}
%
%
\def\eqref#1{(\ref{#1})}
%
%
\def\epspdffile#1{\leavevmode\ifpdf\epsffile{#1.pdf}\else\epsffile{#1.eps}\fi}
%
%
\newif\ifdraft
\draftfalse
\def\note[#1]#2{\message{(#1)}\ifdraft{\noindent\em[#2]\/}\fi}
%
%
\ifdraft
\fi
%
%
\def\erf{\mathop{\rm erf}}          
%
%
\def\rational#1#2{{\mathchoice{\textstyle{#1\over#2}}%
  {\scriptstyle{#1\over#2}}{\scriptscriptstyle{#1\over#2}}{#1/#2}}}
\def\half{\rational12}                      
\def\third{\rational13}                     
\def\quarter{\rational14}                   
\def\O{{\cal O}}                            
\def\defn{\equiv}                           
\def\opt{{\mbox{\tiny opt}}}                
\def\implies{\Rightarrow}                   
%
%
\def\N{{\bb N}}                     
\def\R{{\bb R}}                     
\title{Monte Carlo Integration with Subtraction}

\author{Rudy Arthur\thanks{R.Arthur@sms.ed.ac.uk}
  \hskip1.3em and\hskip1.3em
  A.~D.~Kennedy\thanks{adk@ph.ed.ac.uk} \\[1ex]
  School of Physics and Astronomy, \\
  The University of Edinburgh, The King's Buildings, \\
  Edinburgh, EH9~3JZ, Scotland}
  
\date{\small{\it\today}}

\maketitle

\begin{abstract}
  This paper investigates a class of algorithms for numerical
  integration of a function in \(d\) dimensions over a compact domain by Monte Carlo methods.  We
  construct a histogram approximation to the function using a partition of the
  integration domain into a set of bins specified by some parameters.  We then
  consider two adaptations; the first is to subtract the histogram
  approximation, whose integral we may easily evaluate explicitly, from the
  function and integrate the difference using Monte Carlo; the second is to
  modify the bin parameters in order to make the variance of the Monte Carlo
  estimate of the integral the same for all bins.  This allows us to use
  Student's \(t\)-test as a trigger for rebinning, which we claim is more
  stable than the \(\chi^2\) test that is commonly used for this purpose.  We
  provide a program that we have used to study the algorithm for the case
  where the histogram is represented as a product of one-dimensional
  histograms.  We discuss the assumptions and approximations made, as well as
  giving a pedagogical discussion of the myriad ways in which the results of
  any such Monte Carlo integration program can be misleading.
\end{abstract}

\section{Introduction}

We are interested in evaluating the integral of a function \(f:\mathbb{R}^d\to
\mathbb{R}\) over a compact domain.  It is a simple matter to map any compact
domain into the unit hypercube, so we need to evaluate
\begin{displaymath}
  I = \int_{[0,1]^d} dx\,f(x)=\int_0^1 dx_1\cdots\int_0^1 dx_d\,f(x_1,\ldots,x_d).
\end{displaymath}
\(I\) is just the average value of \(f\) over a uniform probability
distribution which vanishes outside the unit hypercube, we denote this average value by
\(\langle f\rangle\).

Defining the sample average \(\bar f\) over a set of \(N\) uniformly
distributed random points \(\{x^{(i)}\in[0,1]^d, i=1,\ldots,d\}\) to be
\begin{equation}
  \bar f \defn \frac1N \sum_{i=1}^N f(x^{(i)})
  \label{fbar}
\end{equation}
the weak law of large numbers~\cite{Durr} allows the identification \(\langle
f\rangle = \lim_{N\to\infty} \bar f\) assuming only that the integral exists.
Strictly speaking the weak law of large numbers states that, with probability
arbitrarily close to one, \(\bar f\) will become arbitrarily close to \(\langle f\rangle\) 
for sufficiently large~\(N\).

The central limit theorem makes the stronger statement that the probability
distribution of \(\bar f\) tends to a Gaussian with mean \(\langle f\rangle\)
and variance \(V/N\),
\begin{equation}
  \langle f\rangle = \bar f + \O\left(\sqrt\frac VN\right)
  \label{Idef}
\end{equation}
where the variance of the distribution of \(f\) values is
\begin{displaymath}
  V \defn \left\langle\Bigl(f-\langle f\rangle\Bigr)^2\right\rangle
    = \int_{[0,1]^d} dx\,\Bigl(f(x) - \langle f\rangle\Bigr)^2
    = \langle f^2\rangle - \langle f\rangle^2,
\end{displaymath}
but it requires stronger assumptions that we shall discuss shortly.  An unbiased
estimate of the variance is given by
\begin{displaymath}
  \hat V \defn \frac{\overline{f^2} - \bar f^2}{N-1},
  \qquad \langle\hat V\rangle = V,
\end{displaymath}
where of course
\begin{displaymath}
  \overline{f^2} = \frac1N\sum_{i=1}^N f(x^{(i)})^2.
\end{displaymath}
The estimate of the integral \(\bar f\) is within one standard deviation
\(\sigma\defn\sqrt{V/N}\) of the true value \(I=\langle f\rangle\) about 68\%
of the time.

Note that the error is proportional to \(1/\sqrt N\) independent of the
dimension \(d\) of the integral.  From this fact stems the great utility of
Monte Carlo for integration in many dimensions compared to numerical
quadrature.  Generally for numerical quadrature (trapezoid rule, Simpson's
rule etc.) the error is \(\O(\Delta^k)\) where \(\Delta\) is the grid spacing
and \(k\) is a small number.  With a fixed budget of function evaluations,
\(N\), on a regular grid each axis must be divided into \(\root d\of N\)
segments.  So \(\Delta\propto N^{-1/d}\) and thus the error is
\(\O(N^{-k/d})\); therefore in dimension \(d>2k\) the Monte Carlo error is
smaller.  An intuitive explanation for this is that a random sample is more
homogeneous than a regular grid~\cite{James:1980yn}.

\subsection{Singular Integrands}

If the integrand has a singularity within or on the boundary of the integration
region extra care is required.  Let us consider the proof of the central limit
theorem.  The probability distribution \(P_f\) for the values \(F=f(x)\) when
\(x\) is chosen from the distribution \(P\)~is
\begin{equation}
  P_f(F) \defn \int dx\,P(x)\, \delta\Bigl(F-f(x)\Bigr)
  \label{PX}
\end{equation}
for which
\begin{displaymath}
  \int dF\,P_f(F) = \int dx\,P(x) = 1
  \qquad\mbox{and}\qquad
  \int dF\,P_f(F)\,F  = \int dx\,P(x)\,f(x) = \langle f\rangle.
\end{displaymath}
We define the generating function for connected moments as the logarithm of the
Fourier transform of~\(P_f\)
\begin{equation}
  W_f(ik) \defn \ln\int dF\,P_f(F)e^{ikF} 
  = \ln\int dx\,P(x)e^{ikf(x)}
  = \ln\langle e^{ikf}\rangle.
  \label{Womega}
\end{equation}
Assuming that \eqref{Womega} can be expanded in an asymptotic series
\begin{equation}
  W_f(k) = \sum_{m=1}^\infty \frac{k^mC_m}{m!}
  \label{Wseries}
\end{equation}
where coefficients \(C_m\) (cumulants)
\begin{eqnarray*}
  C_0 &=& 1 \\
  C_1 &=& \langle f\rangle \\
  C_2 &=& V = \left\langle\Bigl(f-\langle f\rangle\Bigr)^2\right\rangle \\
  C_3 &=& \left\langle\Bigl(f-\langle f\rangle\Bigr)^3\right\rangle \\
  C_4 &=& \left\langle\Bigl(f-\langle f\rangle\Bigr)^4\right\rangle - 3C_2 \\
  C_5 &=& \left\langle\Bigl(f-\langle f\rangle\Bigr)^5\right\rangle - 10C_3C_2 \\
  C_6 &=& \left\langle\Bigl(f-\langle f\rangle\Bigr)^6\right\rangle
    - 15C_4C_2 - 10C_3^2 - 15C_2^2 \\
  &\vdots&
\end{eqnarray*}
are all finite.  We consider the distribution function \(P_{\bar f}\) for the
sample average \(\bar f\) defined in equation~\eqref{fbar}
\begin{displaymath}
  P_{\bar f}(F) = \int dx^{(1)}\cdots dx^{(N)}\,P(x^{(1)})\cdots P(x^{(N)})\,
    \delta\left(F - \frac1N\sum_{i=1}^N f(x^{(i)})\right) 
\end{displaymath}
and the corresponding generating function \(W_{\bar f}\)
\begin{equation}
  W_{\bar f}(k) \defn \ln\int dF\,P_{\bar f}(F)\,e^{ikF}.
  \label{Wbaromega}
\end{equation}
Since the points \(x^{(i)}\) were chosen independently \eqref{Wbaromega}
factorises to give
\begin{displaymath}
  W_{\bar f}(k) = \ln\left[\int dx\,P(x)e^{ikf(x)/N}\right]^N
    = NW_f\left(\frac kN\right),
\end{displaymath}
expanding which gives an (asymptotic) expansion in powers of \(1/N\)
\begin{displaymath}
  W_{\bar f}(k)
  = \sum_{m=1}^\infty \frac{k^mC_m}{N^{m-1}m!}
  = kC_1 + \frac{k^2C_2}{2N} + \frac{k^3C_3}{6N^2} + \O\left(\frac1{N^3}\right).
\end{displaymath}
Ignoring terms of \(\O(N^{-2})\) and taking the inverse Fourier transform gives
\begin{displaymath}
  P_{\bar f}(F) =
  \frac{\exp\left(\frac{-(F-\langle f\rangle)^2}{2V/N}\right)}{\sqrt{2\pi V/N}},
\end{displaymath}
namely a Gaussian with mean \(\langle f\rangle\) and variance~\(V/N\).

\begin{figure}[htp]
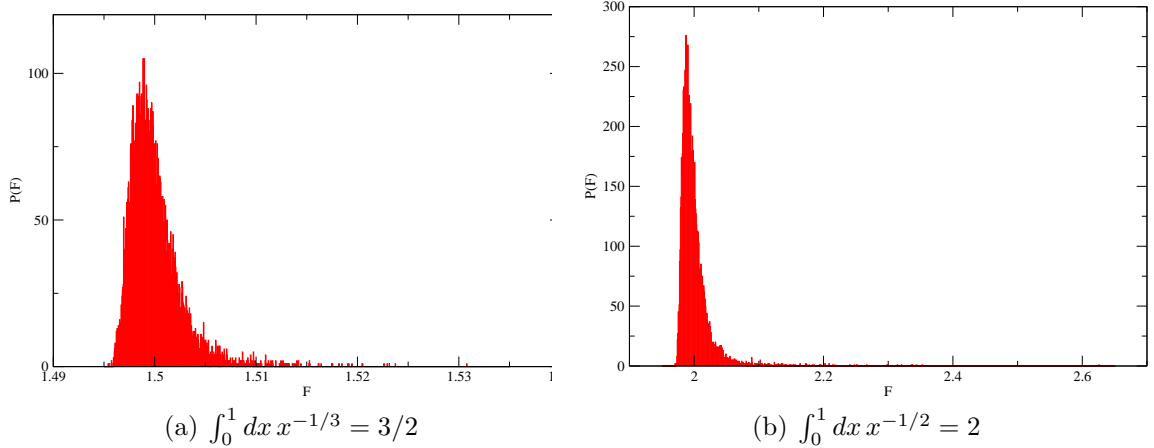

  \begin{center}
    \subfigure[\(\int_0^1 dx\,x^{-1/3}=3/2\)]{%
      \includegraphics[angle=0,width=0.45\textwidth]{x0p3.eps}}
    \subfigure[\(\int_0^1 dx\,x^{-1/2}=2\)]{%
      \includegraphics[angle=0,width=0.45\textwidth]{x0p5.eps}}
  \end{center}
  \caption{(a)~\(\int_0^1 dx\,x^{-\alpha}\) with \(\alpha=\rational13\) and
    (b)~\(\alpha=\half\) evaluated by Monte Carlo with \(10,000\) function
    evaluations.  The distribution of results about the mean is distinctly
    non-Gaussian.  For~(b), which has worse divergences, there is more weight
    in the tail of the distribution. For ~(a) the variance exists but the
    distribution is not Gaussian due to the vanishing of the higher moments.}
  \label{fig:x0.5}
\end{figure}

However, if any of the moments \(C_n\) are not finite then the expansion
\eqref{Wseries} is not justified.  A simple class of integrals with divergent
higher moments is \(\int_0^1 dx\,x^\alpha\) with \(\alpha < 0\).  If
\(-1<\alpha<0\) this integral is well-defined but has an infinite number of
divergent moments.  Following equation~\eqref{PX}
\begin{displaymath}
  P(y) = \int_\epsilon^1 dx\,\frac{\delta(y-x^\alpha)}{1-\epsilon}
  = \frac{ y^{\frac1{\alpha}-1}}{\alpha(1-\epsilon)}
\end{displaymath}
for \(y\in[\epsilon^\alpha,1]\), and zero elsewhere.  The cumulants are
combinations of the moments
\begin{displaymath}
  \int dy\,P(y) y^n = \frac{1-\epsilon^{1+\alpha n}}{(1-\epsilon)(1+n\alpha)}
\end{displaymath}
which diverge as \(\epsilon\to0\) if \(1+\alpha n\leq0\), or equivalently
every cumulant \(C_n\) with \(n>-1/\alpha\) diverges.  What this means in
practice is that although estimating such integrals by Monte Carlo is allowed
--- the weak law of large numbers assures us that as long as \(N\) is large
enough the estimate will converge but it gives no indication of how large 
\(N\) should be --- it is misleading to estimate the error from the
variance alone. This is because the distribution of the estimates of the integral is not
Gaussian even if the variance exists.  In practice one obtains non-Gaussian
distributions with ``fat tails'', for some examples see Figure \ref{fig:x0.5}.

We may also observe that a singularity in the integrand does not necessarily
lead to infinite cumulants: for the function \(f(x)=-\ln x\) a similar analysis
to that given above shows that \(P_f(F)=e^{-F}\) for \(F\in[0,\infty)\) and
hence \(\langle f^m\rangle=m!\) so all its cumulants are finite, and therefore
the Monte Carlo estimates of the integral do have a Gaussian distribution as
the number of samples \(N\to\infty\).

For all convergent integrals Monte Carlo provides an estimate of the mean.
However, if any moment diverges the distribution of the means is not Gaussian,
and in general even if the variance exists it gives an underestimate of the
``width'' of the probability distribution.  This is often the case in
practice, for example when evaluating Feynman parameter integrals which are
integrable but not square integrable.  If the Monte Carlo integrator is
treated as a black box the quoted error from the standard deviation will often
be an underestimate of the true error.  With these provisos on the
applicability of Monte Carlo integration we now turn to our main topic, a new
method of adaptive Monte Carlo integration.

\section{Variance Reduction}

The Monte Carlo scheme of Section 1 (na\"\i ve Monte Carlo) can be improved by
variance reduction schemes~\cite{Ham}: the basic idea is to use some
information about the integral in order to reduce the variance sample average.
We describe two methods: importance sampling and subtraction.

\subsection{Importance Sampling} \label{sec:is}

Let \(\rho(x)\) be a probability distribution, normalised to one, that closely
approximates our function over the interval.  If we generate random points \(x\)
chosen from the distribution \(\rho\) and construct the average \(\langle
f(x)/\rho(x)\rangle\) over these points then we have an estimate of the integral
\begin{equation}
  I = \int dx\,f(x) = \int\rho(x)dx\,\frac{f(x)}{\rho(x)}
  \label{impdef}
\end{equation}
and the variance
\begin{equation}
  V(\rho) \defn \int \rho(x)dx\, \left(\frac{f(x)}{\rho(x)} - I\right)^2.
  \label{impvdef}
\end{equation}
The value \(I\) of \eqref{impdef} does not depend on \(\rho\), but the
corresponding variance of \eqref{impvdef} does, so we may minimise the variance
\eqref{impvdef} by varying the function \(\rho\) subject to the constraint that
it is correctly normalised \(N(\rho)\defn\int dx\,\rho(x)=1\).  Performing the
required functional differentiation with the Lagrange multiplier~\(\lambda\)
\begin{displaymath}
  \frac{\delta[V(\rho)+\lambda N(\rho)]}{\delta\rho(x)}=0
\end{displaymath}
we obtain the result
\begin{displaymath}
  \rho_\opt(x) = \frac{|f(x)|}{\int dx'\,|f(x')|}.
\end{displaymath}
The corresponding value for the minimal variance is
\begin{equation}
  V_\opt \defn V(\rho_\opt) 
    = \left(\int dx\,|f(x)|\right)^2 - \left(\int dx\,f(x)\right)^2,
  \label{impvopt}
\end{equation}
which vanishes if \(f\) has the same sign everywhere in the domain of
integration.

\subsection{Subtraction}

In this case we have
\begin{equation}
  I = \int dx\,f(x) = \int dx\,[f(x)-g(x)] + \int dx\,g(x),
  \label{condef}
\end{equation}
where the integral of \(g\) is known exactly: the corresponding variance is
\begin{equation}
  V(g) \defn \int dx\,[f(x)-g(x)]^2.
  \label{convdef}
\end{equation}
As before the integral \(I\) of \eqref{condef} is independent of \(g\) whereas
the variance is not, so we may minimise the latter by varying~\(g\).  This
time there is no constraint to be imposed\footnote{In practice~\(g\) will be a
  piecewise constant approximation to~\(f\), but we do not impose this as a
  constraint.} and we find
\begin{displaymath}
  \frac{\delta V(g)}{\delta g(x)} = 0 \implies g_\opt = f,
\end{displaymath}
and the optimal variance \(V_\opt=V(g_\opt)=0\).  We therefore choose \(g(x)\)
such that \(g(x)\) is a good approximation to~\(f(x)\).

Both of these methods assume knowledge of the function that one may not have
for complicated multidimensional integrals that arise in practice, thus it is
necessary to find a way to construct \(\rho\) or \(g\) from a function treated
as a black box.

\section{Adaptive Algorithms}

The usual way to construct \(\rho(x)\) is the method of \cite{Lepage:1977sw},
VEGAS.  The key feature of this algorithm is that it is adaptive, that is, it
automatically constructs a probability density that approximates \(|f(x)|\).

Our analysis of the variance reduction in \S\ref{sec:is} assumed that the
function is representable as a product of functions in each variable, although
the correctness of the global Monte Carlo estimate of the integral does not
depend upon this.  Such a factorisation depends on the co-ordinate system, see
\cite{Ohl:1998jn}: in the following we will always work in Cartesian
co-ordinates and assume that the function approximately factorises.

Each axis is divided into a number of bins and we select a point with
each coordinate lying equiprobably in any bin along the corresponding axis;
the points are thus chosen to lie in equiprobably in any of the boxes that are
defined by the intersection of the bins.  The probability distribution
\(\rho\) in \eqref{impdef} is \(\rho(x) = 1/|b_x|\) where \(|b_x|\) is the
volume of the box \(b_x\) containing~\(x\).  To converge to the optimal grid
the bins are resized so that each bin makes an equal contribution to the
integral of $|f(x)|$.

The analysis in \cite{Lepage:1977sw} assumes that the probability distribution
\(\rho(x)=\prod_{i=1}^d \rho_i(x_i)\) factorises but does not explicitly
assume that the integrand \(f\) does.  In this case the variance is
\begin{displaymath}
  V(\rho) = \int\left(\prod_{i=1}^d \frac{dx_i}{\rho_i(x_i)}\right) f(x) - I^2,
\end{displaymath}
and minimisation with respect to any particular \(\rho_i\) subject to the
normalisation constraint \(N(\rho_i) = \int dx_i\,\rho_i(x_i) = 1\) gives
\begin{displaymath}
  \rho_{i,\opt}(x_i) \propto
  \sqrt{\int\left(\prod_{j=1\atop j\neq i}^d\frac{dx_j}{\rho_j(x_j)}\right)f(x)^2}.
\end{displaymath}
This gives the optimal solution for \(\rho_i\) when all the other
\(\rho_j\) are fixed, but does not immediately give the optimal solution for
all the factors of \(\rho\) simultaneously.  If we make the further assumption
that \(f(x)=\prod_{i=1}^d f_i(x_i)\) then the solution reduces to that which
we found in \S\ref{sec:is}, namely \(\rho_{i,\opt}(x_i)\propto|f_i(x_i)|\) and
hence \(\rho_\opt(x)\propto|f(x)|\).

We may identify some issues with this approach.  Consider the optimal
distribution \(\rho_\opt\) that VEGAS is trying to find.
\(V_\opt\) of equation \eqref{impvopt} is not necessarily zero unless \(f(x)\) has the
same sign everywhere.  A simple example where \(V_\opt\neq0\) is in one
dimension with \(f(x) = \sin2\pi x\), which gives \(V_\opt=16\).  In simple
cases it is possible to divide up the integration region into parts in which
the sign of the integrand does not change, however this assumes detailed
knowledge of the function and may not be feasible in a high number of
dimensions.  Compare this to the subtraction method, for which \(V_\opt=0\) in
this case.

Secondly, there seems to be no automatic method for deciding when VEGAS has
converged to a (near) optimal grid.  Usually it is advised to use as few
function evaluations as possible until the grid approximately converges and
then sample on the optimal grid.  To test this one computes the \(\chi^2\)
statistic.  If on each iteration of VEGAS an independent estimate of the
integral \(I_i\) is produced using \(N'\) function evaluations; after \(n\)
iterations the variables \((I_1-I), \ldots, (I_n-I)\) will have a normal
distribution with mean zero and variance \(\sigma^2\), assuming that \(N'\) is
sufficiently large for central limit theorem to apply.  Therefore the sum of
their squares
\begin{equation}
  \chi^2 = \sum_{i=1}^n \left(\frac{I_i-I}\sigma\right)^2;
  \label{chisq}
\end{equation}
has a \(\chi^2\) distribution with \(n\) degrees of freedom
\begin{displaymath}
  P^{\chi^2}_n(\chi^2)
    = \frac{(\chi^2)^{\frac n2-1}e^{-\chi^2/2}}{2^{n/2}\Gamma\left(\frac\nu2\right)}
\end{displaymath}
for \(\chi^2\geq0\).  Sadly, we cannot compute this \(\chi^2\) since it
requires the \emph{exact} values of \(I\) and \(\sigma\).  If we use the
estimate \(\bar I\approx I\) then it is easy to show that the resulting
quantity
\begin{displaymath}
  \chi'^2 = \sum_{i=1}^n \left(\frac{I_i-\bar I}\sigma\right)^2
\end{displaymath}
still has a \(\chi^2\) distribution but now with \(n-1\) degrees of freedom.
However, since \(\sigma^2\) occurs in the denominator of \(\chi^2\) attempts to
replace it with some stochastic estimate will tend to give anomalous large
values for \(\chi^2\) or, to make a more precise statement, the distribution
will not be a \(\chi^2\) distribution but something with fatter tails.

Finally, we observe that the VEGAS rebinning algorithm 
will tend to produce small bins where the function value is large.  
If the integrand has a relatively flat but high plateau with steep edges this
approach will tend to calculate the integral well in the flat region, where it
is easy, and miss the regions of large variance which require more function
evaluations to estimate accurately.  We will see an explicit example in
section~\ref{sec:numex}.

\section{Our Algorithm}

Our algorithm uses two adaptations, subtraction and rebinning, and uses
Student's \(t\)-test as a robust trigger to decide when to apply them.

As in all such adaptive algorithms, we trigger an adaptation whenever there is
statistically significant evidence that the current histogram parameters are
not optimal, and the na\"\i ve goal is that the histogram parameters will
converge to their optimal values thereby minimising the variance of our Monte
Carlo estimate of the integral.  We say ``na\"\i ve'' because this clearly
cannot happen: we are constructing an ergodic Markov process and such a
process must converge to a fixed point distribution of histogram parameters
which is non-zero everywhere --- it cannot converge to a single point in the
parameter space.  Of course, we may hope that the Markov process will converge
to a distribution of histogram parameters strongly peaked about the optimal
ones, but how well this may be achieved depends in a complicated way upon lots
of the algorithmic details and approximations: for example, how many samples
we have to take in each bin before using the central limit theorem to assert
that their mean is normally-distributed, or how much we choose to damp the
rebinning adaptations to try to avoid instabilities.

We should put such worries in context, it is easy to see that for \emph{any}
adaptive Monte Carlo integration scheme there are functions for which they
will give an answer with an unreliable estimate of its error.  A simple
example is an approximation to a \(\delta\) function within a region where the
function is zero: not only will Monte Carlo algorithms not ``see'' the
\(\delta\) function, but adaptive importance sampling will make it less likely
for them to do so.

\subsection{Subtraction Adaptation} \label{sec:sub}

We construct an approximation function \(\hat f\) adaptively from the Monte
Carlo process.  Like \cite{Lepage:1977sw} we assume that the integrand may be
reasonably approximated by a product of one-dimensional histograms with \(n\)
bins along each axis.  This means we only have to accumulate \(\O(nd)\) rather
than \(\O(n^d)\) values, and more significantly we can afford to evaluate a
reasonably large number of samples per bin whereas it would be impossible to
evalute even one sample per box in practice for \(n=10\) in \(d=10\) dimensions
for example.  However, we stress that the this is not a requirement of our
method: the idea of using adaptive subtractions as well as adaptive importance
sampling is independent of the choice of histogram representation.  Subtraction
could be used together with recursive stratified sampling \cite{Press:MISER}
for example, and the function approximation stored not as a product of
histograms as we will describe below but as independent values in each box
as required.

We should also stress that the Monte Carlo algorithm will give the correct
value for the integral even if the integrand is not well approximated by a
product of histograms, or even as a product at all.  All that happens in such
cases is that we have no good reasons to expect our adaptations to
significantly reduce the statistical error in the result.

We describe a method closely following VEGAS, assuming that our function is
representable as a product of one-dimensional functions.  Each axis is divided
into a number of bins, with a bin along an axis defined to be the Cartesian
product of an interval along that axis and the unit interval along every other
axis.  The intersection of \(d\) bins, one along each axis, will be called a
box.

We generate quasi-random points for the Monte Carlo sampling by choosing a
random permutation of the bins and a random point in each bin.  This ensures
that each bin has an equal number of samples while still sampling the entire
space homogeneously.  This implements the importance sampling distribution
\(\rho\), as the probability of selecting a sample point \(x\) within any box
\(b\) is \(1/n^d\), and therefore the probability density \(\rho(x)=1/V_b\)
where \(V_b\) is the volume of the box~\(b\).  Initially we choose all the
histogram bins to have equal width, and therefore all the boxes to have the
same volume and therefore \(\rho(x)\) equal everywhere, for want of any better
information.  Likewise, we initially set the subtraction function \(\hat f(x)=0\)
everywhere.

We generate a set \(\{x^{(i)},\ldots,x^{(N)}\}\) of quasi-random points chosen
from the distribution \(\rho\), and for each bin \(\beta_\mu\in\{1,\ldots,n\}\)
along each axis \(\mu\in\{1,\ldots,d\}\) we accumulate the quantities
\begin{displaymath}
  \Sigma_k(\beta_\mu) 
    = \sum_{x^{(i)}\in\beta_\mu} \left(\frac{f(x^{(i)})}{\rho(x^{(i)})}\right)^k
\end{displaymath}
for \(k\in\{0,1,2\}\).  \(\Sigma_0(\beta_\mu)\) is just the number of samples
in the bin \(\beta_\mu\), so with our quasi-random number strategy for
generating sample points it is just equal to the integer \(N/n\) for all bins,
so we do not really need to accumulate it, although we might if we used a
different random point generator.  We define
\begin{displaymath}
  \hat f_{\beta_\mu} \defn \frac{\Sigma_1(\beta_\mu)}{\Sigma_0(\beta_\mu)}
  \approx \int_{x\in\beta_\mu} \rho(x)dx\,\frac{f(x)}{\rho(x)}
    = \int_{x\in\beta_\mu} dx\,f(x)
\end{displaymath}
where \(x^{(i)}\in\beta_\mu\) means that \(x^{(i)}\) falls within \(\beta_\mu\),
that is the \(\mu\) coordinate \(x^{(i)}_\mu\) of the point \(x^{(i)}\) lies in
the interval \(\beta_\mu\) along the \(\mu\) axis.  

If \(f\) is a product of one-dimensional functions, \(f(x)=f_1(x_1)\cdots
f_d(x_d)\) then
\begin{displaymath}
  \int_{x\in\beta_\mu} dx\,f(x) = \int_{\beta_\mu} dx_\mu\,f_\mu(x_\mu)
      \prod_{\nu\neq\mu} \int_0^1 dx_\nu\,f_\nu(x_\nu);
\end{displaymath}
moreover, if the average value of \(f_\mu\) in the bin \(\beta_\mu\) is
\begin{displaymath}
  \langle f_\mu\rangle_{\beta_\mu}
    = \frac1{|\beta_\mu|} \int_{\beta_\mu} dx_\mu\,f_\mu(x_\mu)
\end{displaymath}
with \(|\beta_\mu| = \int_{\beta_\mu} dx_\mu\) being the bin width, then
\begin{displaymath}
  \hat f_{\beta_\mu} \approx \int_{x\in\beta_\mu} dx\,f(x)
    = \frac{\langle f_\mu\rangle_{\beta_\mu} |\beta_\mu| I}
      {\displaystyle\int_0^1 dx_\mu\,f_\mu(x_\mu)}
\end{displaymath}
where
\begin{displaymath}
  I = \int dx\,f(x) = \prod_{\nu=1}^d \int_0^1 dx_\nu\,f_\nu(x_\nu)
\end{displaymath}
is the integral we wish to evaluate.  We thus may construct an estimate of the
value of the function \(f(x)\) as
\begin{equation}
  \hat f(x) 
    = \prod_{\nu=1}^d \frac{\hat f_{\beta_\nu}}{|\beta_\nu| \hat I^{d-1}}
    = \frac1{V_b\hat I^{d-1}} \prod_{\nu=1}^d \hat f_{\beta_\nu}
    \approx \prod_{\nu=1}^d \langle f_\nu(x_\nu)\rangle_{\beta_\nu}
  \label{norm}
\end{equation}
where \(x\in\beta_\nu\) for \(\nu\in\{1,\ldots,d\}\), \(V_b = |\beta_1|\cdots
|\beta_d|\) is the volume of the box \(b\) containing~\(x\), and
\begin{displaymath}
  \hat I = \frac1N \sum_{i=1}^N \frac{f(x^{(i)})}{\rho(x^{(i)})}
    \approx I
\end{displaymath}
is the current ``global'' estimator for the integral.

We use this function for our subtraction adaptation, that is we take \(g=\hat
f\) in~\eqref{condef}, and evaluate
\begin{equation}
  I' \defn \int dx\,[f(x)-\hat f(x)]
    = \int \rho(x)dx\,\frac{f(x)-\hat f(x)}{\rho(x)}
  \label{eqn:iprime}
\end{equation}
by Monte Carlo. The integral of \(\hat f\) is known exactly
\begin{displaymath}
  I'' \defn \int dx\,\hat f(x)
  = \frac1{\hat I^{d-1}} \prod_{\nu=1}^d \sum_{\beta_\nu=1}^n \hat f_{\beta_\nu},
\end{displaymath}
whence we can use \eqref{condef} and set \(I=I'+I''\).

\subsection{Rebinning Adaptation}

After making a subtraction as described in section~\ref{sec:sub} the integrand
is everywhere approximately zero to the best of our current knowledge, so the
usual importance sampling analysis of section~\ref{sec:is} does not give us any
useful way of choosing the function~\(\rho\).  We may therefore choose \(\rho\)
so as to make the variance constant over all boxes, which in our case
corresponds to making it constant for all bins along each axis.  We introduce
an estimator for the variance with the bin \(\beta_\mu\)
\begin{displaymath}
  \hat V_{\beta_\mu}
  \defn \frac{\Sigma_0(\beta_\mu)}{\Sigma_0(\beta_\mu)-1}
    \left(\frac{\Sigma_2(\beta_\mu)}{\Sigma_0(\beta_\mu)}
      -\hat f_{\beta_\mu}^2\right)
  \approx \int_{x\in\beta_\mu} \rho(x)dx\,\left(\frac{f(x)}{\rho(x)} - I\right)^2,
\end{displaymath}
and our rebinning step adjusts the bin parameters to make the integral of this
histogram constant within each new bin.  An interpolation is performed along
each axis to get the value of the approximation function in the new bins.  The
interpolation uses cubic splines to fit the current function approximation in
the current set of bins: cubic splines perform better for some of our test
integrals than lower-order interpolation, although there is no intrinsic
reason to prefer them.  When the new set of bins have been found the value of
the interpolation in the centre of that bin is used for $\hat
f_{\beta_\mu}$. As with importance sampling it is possible that the algorithm
will rebin too much and narrow the bins very strongly about regions of high
variance, thus we damp the rebinning algorithm by adding a small constant
variance to each bin so that no bin can have width zero. The second part of
~\eqref{eqn:iprime} tells us how to make a subtraction even after changing the
importance sampling \(\rho\): we just keep the old bins to evaluate \(\hat f\)
(perhaps packaging them in a closure) while generating points according to the
new bins specified by \(\rho\).  The value of \(\hat f\) for the next
iteration is accumulated in the new bins. Interpolation was easier to use in
our current implementation and allows us to combine estimates from different
iterations and avoid starting again after each rebinning.

\subsection{Student Trigger}

If our two adaptations have achieved their goal then the estimates
\(z_{\beta_\mu}\) of \([f(x)-g(x)]/\rho(x)\) within bin \(\beta_\mu\) should
have mean zero and equal variance, and if we average enough samples within each
bin to be able to apply the central limit theorem then they should follow a
Gaussian distribution.  Even though we do not know the exact value of the
variance, we can test this hypothesis without any further approximations by
computing the quantity
\begin{equation}
  t \defn \frac{\mu_z}{\sqrt V_z}
  \label{tstat}
\end{equation}
where
\begin{displaymath}
  \mu_z \defn \frac1{nd} \sum_{\mu=1}^d \sum_{\beta_\mu=1}^n z_{\beta_\mu}
\end{displaymath}
is the average of \(z\) over all bins and
\begin{displaymath}
  V_z \defn \frac1{nd(nd-1)} \sum_{\mu=1}^d \sum_{\beta_\mu=1}^n (z_{\beta_\mu} - \mu)^2
\end{displaymath}
an unbiased estimate of the variance of \(\mu_z\).  \(t\)~has a Student
\(t\)-distribution with \(nd-1\) degrees of freedom, where Student's
distribution with \(\nu\) degrees of freedom is
\begin{displaymath}
  P^t_\nu(t) = \frac{\left(1 + \frac{t^2}{\nu}\right)^{-\frac{\nu+1}2}}
    {\sqrt\nu B\left(\frac\nu2,\frac12\right)}.
\end{displaymath}

Let \(k = A^{-1}(p,nd-1)\) where \(A^{-1}\) is the inverse cumulative Student
distribution, so we expect that \(-k\leq t\leq k\) with probability \(p\).
Therefore, if \(|t|>k\) we may exclude the hypothesis that we have Gaussian
distributed bin values with mean zero and the same variance with probability
\(p\) and carry out a new adaptation.  This should give a more stable
condition for convergence towards the optimal approximation function.  Each
axis could have a separate trigger, but there does not seem to be any obvious
advantage, just as there seems to be no reason to have a different
number of bins \(n\) along each axis.

The rebinning algorithm is a Markov process that is hopefully converging to
the ``optimal'' bin distribution.  However, as we stated before, an ergodic
Markov process \emph{cannot} converge to a delta function, so our automated
rebinning procedure cannot find the optimal integration parameters (the bin
widths and subtraction values).  In practice this manifests itself as an
instability in the rebinning algorithm.  This is not a problem specific to our
method but is shared by all automatic rebinning schemes; we hope and expect
that our use of Student's test as a trigger will be more stable that those
that use a \(\chi^2\) trigger, but to make this into a more quantitative
statement would require at least a significant restriction on the class of
integrands under consideration.

\section{Numerical examples} \label{sec:numex}

\begin{figure}[htp]
  \psfrag{t(x)}{\huge \(t(x)\)}
  \psfrag{x}{\huge \(x\)}
  \begin{center}
    \subfigure[After adaptive subtraction]{%
      \includegraphics[angle=-90,width=0.45\textwidth]{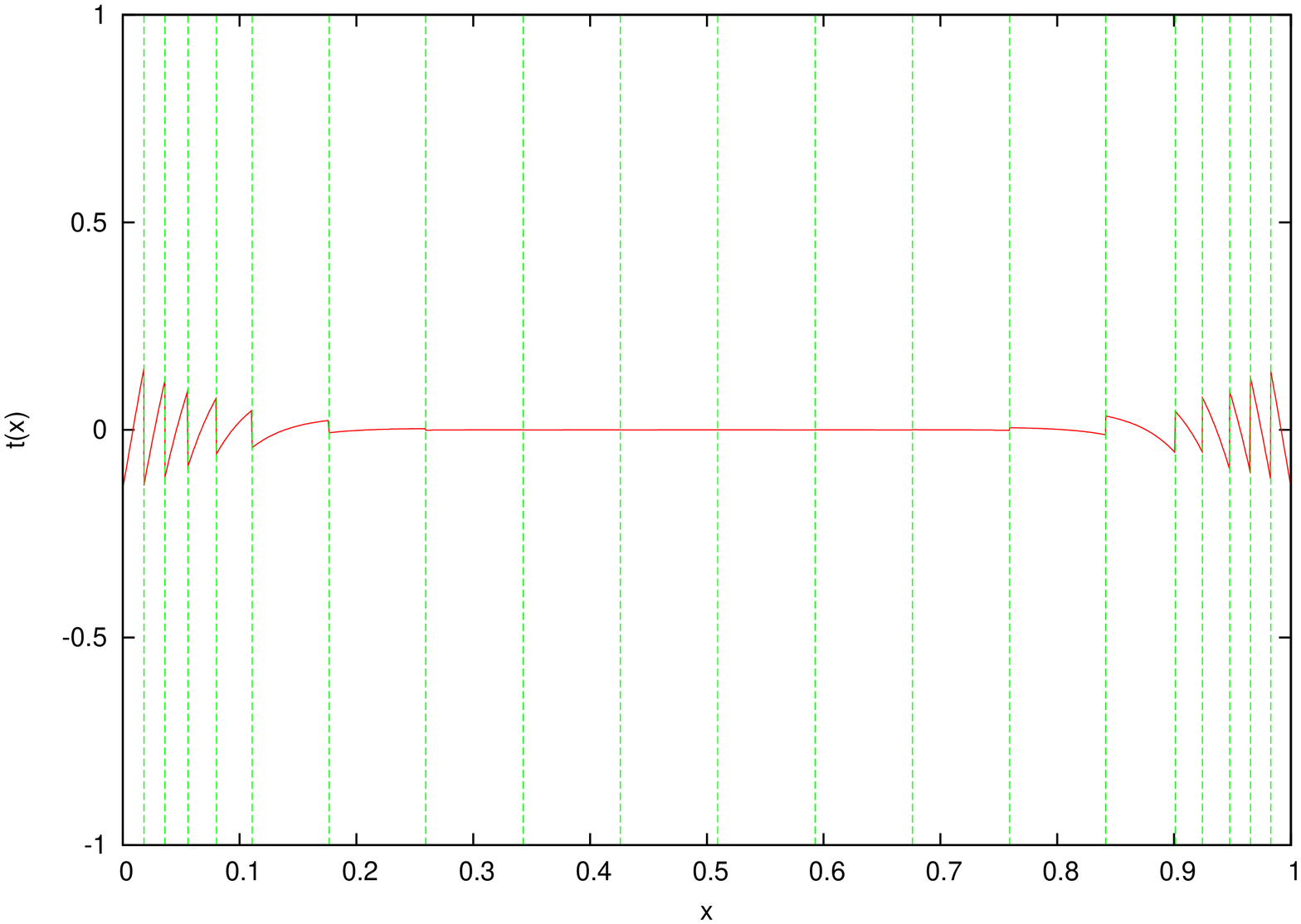}}
    \subfigure[After adaptive importance sampling]{%
      \includegraphics[angle=-90,width=0.45\textwidth]{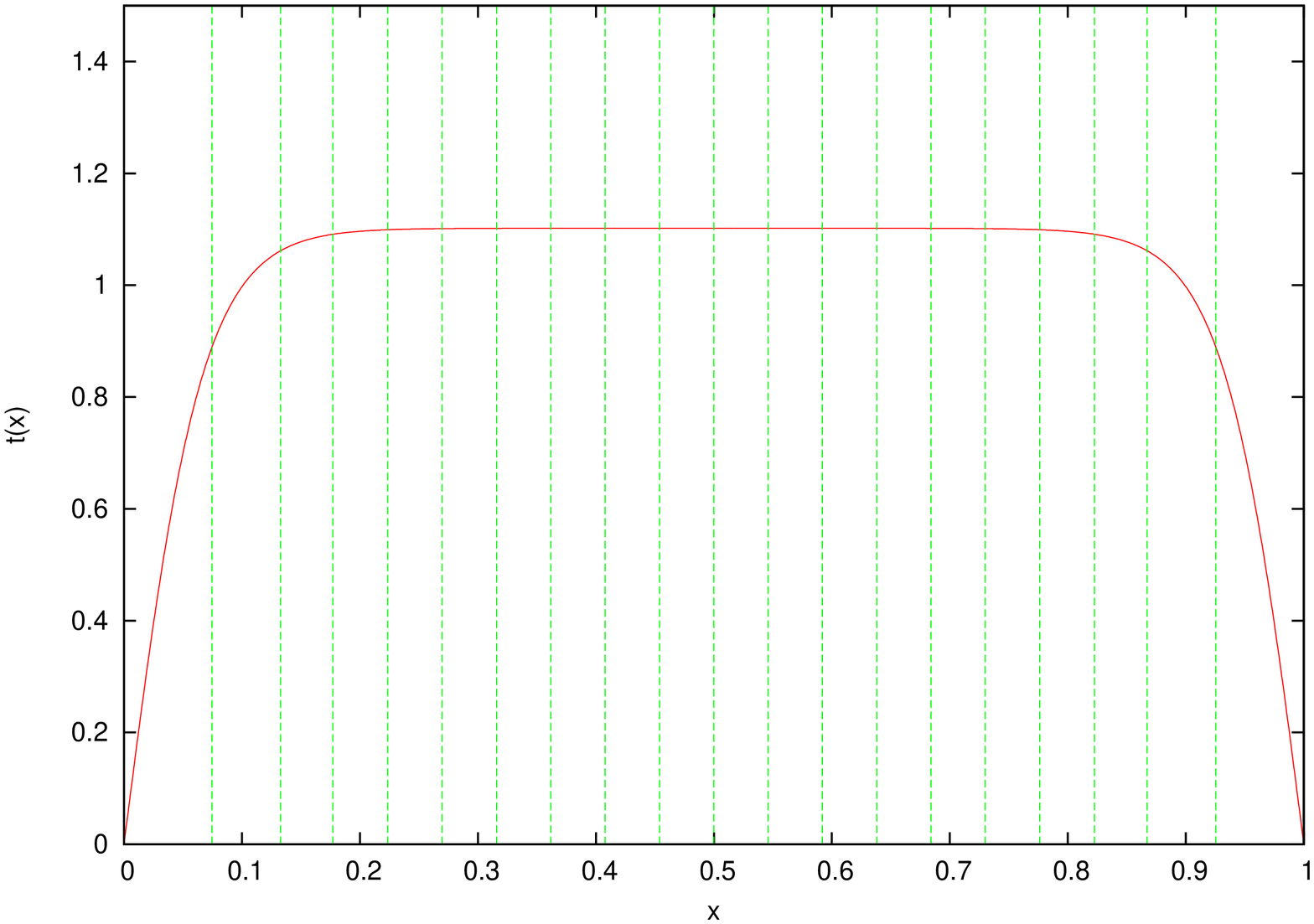}}
    \caption{Rebinning with our algorithm~(a) or importance sampling~(b) for
      the integral \eqref{tst0} along with the function \(t(x)-\hat t(x)\)
      where \(\hat t\) is the approximation function adaptively constructed by
      our algorithm.  The bin distribution for our algorithm concentrates the
      bins at the edges where the variance is largest.}
    \label{fig:edge}
  \end{center}
\end{figure}

\begin{figure}[htp]
  \psfrag{log(s/f)}{\huge \(\log(\frac{s}{ \bar{t}} )\)}
  \psfrag{n}{\huge \(n\)}
  \begin{center}%
    \includegraphics[angle=-90,width=0.8\textwidth]{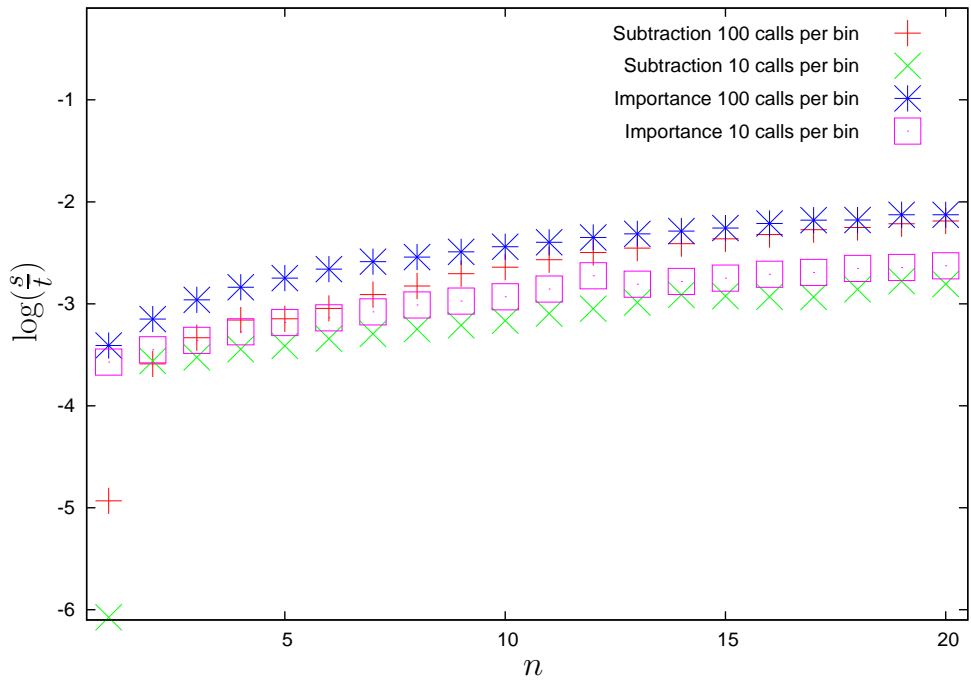}
    \caption{Log of the standard error over the integral evaluation calculated
      as described in the text for equation \eqref{tst1} versus dimension.  In
      one dimension the subtraction method is many orders of magnitude better,
      in more than one dimension the subtraction method is still better than
      importance sampling.  For this integrand, having fewer calls per bin is
      better for both subtraction and importance sampling.}
    \label{fig:tanh_dim}
  \end{center}
\end{figure}

To demonstrate the procedure of adaptive subtraction and how it compares to
adaptive importance sampling, we use the {\tt C} language implementation of
VEGAS importance sampling in the GNU Scientific Library (GSL) for
comparison~\cite{GSL}.  We first choose an example that illustrates the
strengths of our method,
\begin{equation}
  I = \int_0^1 dx\,t(x)
  \qquad\mbox{where}\qquad
  t \defn N\tanh(15x) \tanh\Bigl(15(1-x)\Bigr)
  \label{tst0}
\end{equation}
with \(N\) chosen to make the integral \(I=1\).  The integrand is approximately
constant throughout most of the integration region except at the very edges
where it rises rapidly.  Figure \ref{fig:edge}(a) shows the integrand after
subtraction with the new bins, chosen such that the variance in each is equal.
The integration grid is finest in regions where the function is changing
rapidly, this is as it should be, integrating regions where the integrand is
approximately constant requires far fewer function evaluations.  Compare to
part~(b) of the figure which shows the bin distribution produced by importance
sampling.  The bin distribution has moved to where the function itself is
largest.  We integrate,
\begin{equation}
  \int_{[0,1]^d} dx_1\cdots dx_d\,t(x_1)...t(x_d)
  \label{tst1}
\end{equation}
for various values of \(d\) and plot the relative error in
Figure~\ref{fig:tanh_dim}.  This figure shows the log of the standard error
over the integral approximation using importance sampling and our subtraction
method.  We use \(8 \times 10^4\) evaluations, with the number of bins chosen
in both cases so that there are \(10\) or \(100\) calls per bin.  This gives
four opportunities to recalculate the bin distribution (rebin) where we apply
Student's test to decide.  In almost every case, with \(90\%\) probability only
one rebinning is necessary.  We also check the \(\chi^2\) for the importance
sampling case the same number of times and rebin if the \(\chi^2\) per degree
of freedom of the integral estimates differs from one by more than~\(0.5\).
This tends to rebin on every step.  We could combine all the estimates,
weighted by their standard deviation however this sometimes underestimates the
error.  Instead the number we quote comes from \(2\times10^4\) calls on the
bin distribution calculated by the process above.  Figure~\ref{fig:tanh_dim}
shows for this example our algorithm gives more accurate results.

\begin{figure}[htp]
  \psfrag{tstat}{\huge \(t\)}
  \psfrag{cstat}{\huge \(|\chi^2 - 1|\)}
  \psfrag{chisq}{\large \(|\chi^2 - 1|\)}
  \psfrag{it}{\huge \(Iteration\)}
  \begin{center}
    \subfigure[Subtraction Method \(t\)-statistic]{%
      \includegraphics[angle=-90,width=0.45\textwidth]{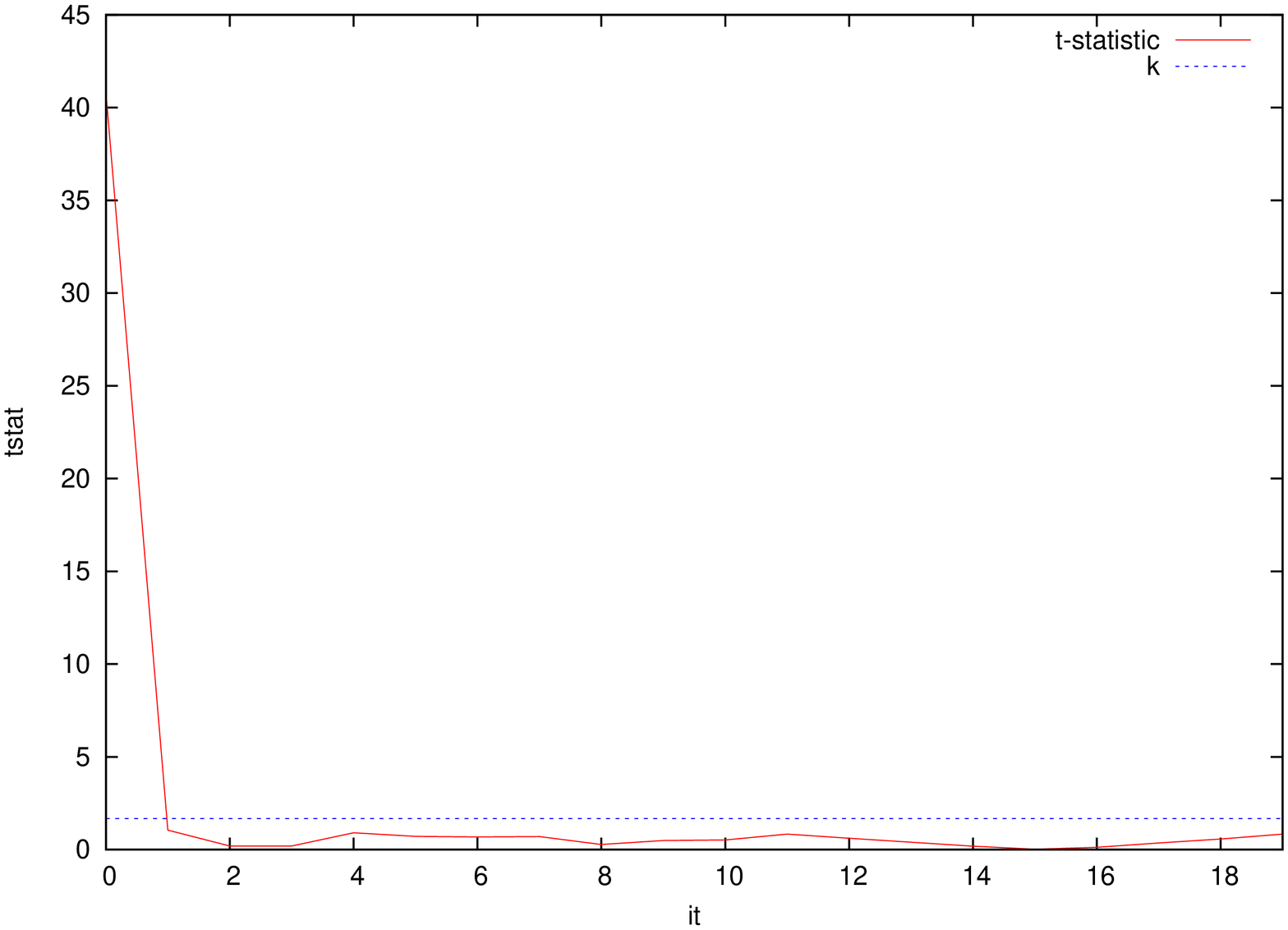}}
    \subfigure[Importance Sampling \(|\chi^2 - 1|\)]{%
      \includegraphics[angle=-90,width=0.45\textwidth]{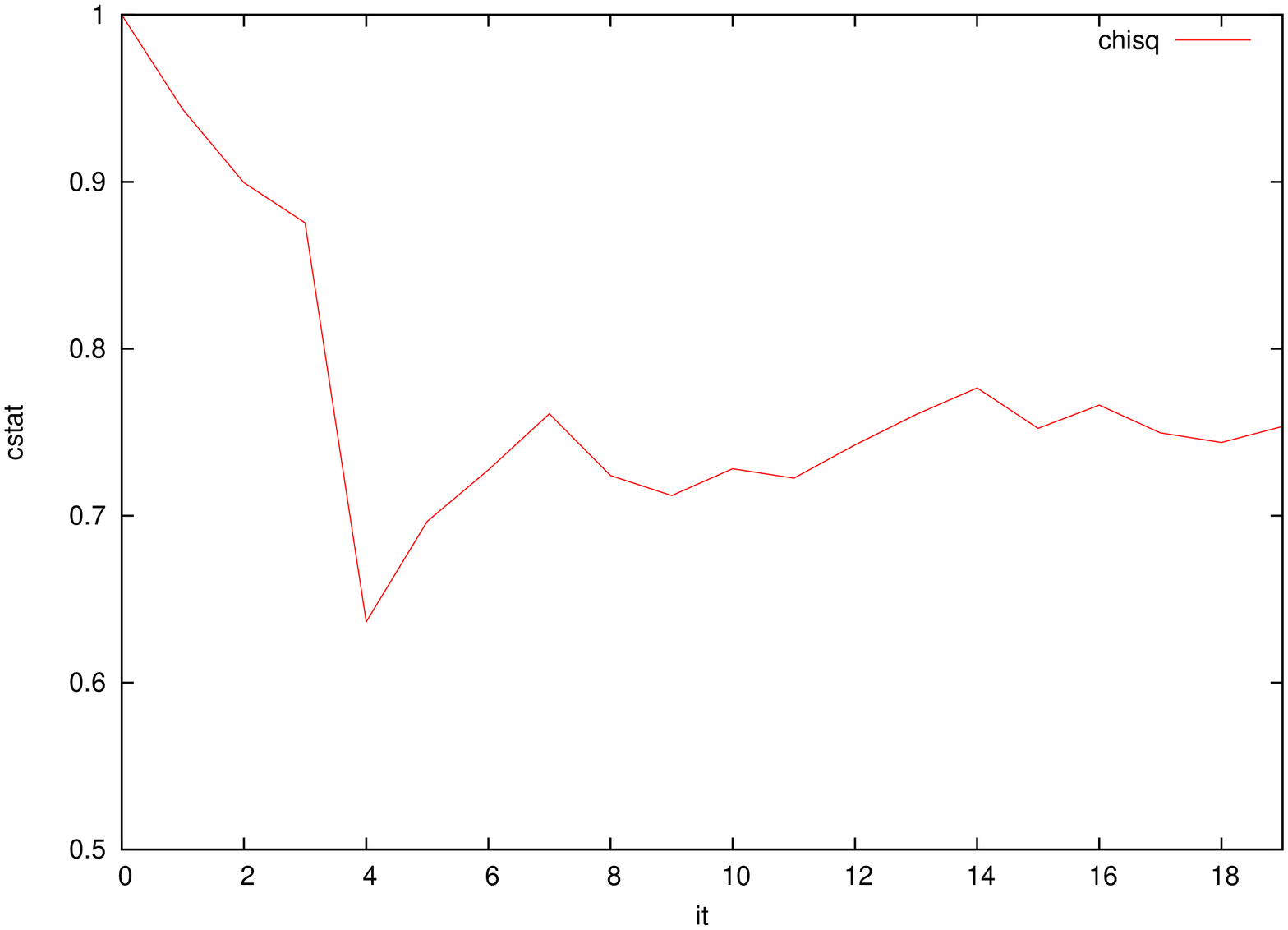}}
    \caption{(a)~The \(t\) statistic \eqref{tstat} (b)~and \(|\chi^2-1|\)
      \eqref{chisq} for the integral \eqref{tst1} in \(4\) dimensions, using
      subtraction~(a) and importance sampling~(b).  Student's \(t\)-test
      triggers once, the \(\chi^2\) test triggers on every iteration.}
    \label{fig:chistu}
  \end{center}
\end{figure}

For the integral \eqref{tst1} in \(4\) dimensions we plot the value of the
\(t\) statistic in figure \ref{fig:chistu}(a), compared
with \(k = A^{-1}(p,n-1)\) for \(p = 0.9\) and \(n=20\), (\(8,000\) calls per
iteration, \(100\) calls per bin) for a number of iterations.  We see that
Student's test triggers only on the first iteration and then not after, the
approximation function is doing as good a job as can be expected.  The closest
equivalent for importance sampling is the \(\chi^2\) per degree of freedom
which we show in figure \ref{fig:chistu}(b).  This is far from \(1\) at every
iteration and the rebinning algorithm triggers at every step.  We could relax
the condition on the \(\chi^2\), e.g., \(|\chi^2-1|<1\) to rebin less often but
this example is indicative of both methods, Student's test triggers relatively
infrequently and the \(\chi^2\) test triggers more frequently, meaning we are
less certain when we have found the optimal bin distribution.

Another test integrand is,
\begin{equation}
  \prod_i^d \sin2\pi x_i.
  \label{tst2}
\end{equation}
In one way this is an excellent test case since the integrand oscillates and
so \eqref{impdef} implies that even with optimal binning importance sampling
cannot reduce the variance to zero.  However this example shows a distressing
feature of our algorithm.  As equation \eqref{norm} shows when calculating an
estimate for the function in each bin we have to divide by the total integral,
if this is zero there will be problems in \(d>1\).  In practice it is unlikely
that we have to integrate a function whose integral is exactly zero.  The root
of this problem is that our knowledge of the integrand is represented as a
product of histograms, so if the integral of the histogram along any axis
vanishes we have no information about how the integrand behaves along the
other axes.

If the value of the subtraction function in every box were stored this would
not be an issue.  We have tried this approach, which is feasible for small
dimensions and small number of bins, on this integrand.  In \(2\) dimensions
with \(20,000\) function evaluations to accumulate the function approximation,
an additional \(20,000\) to evaluate the integral and \(25\) bins per axis we
accumulated the approximation in each of the \(625\) boxes to obtain the
estimates
\begin{eqnarray*}
  \text{Naive} : -0.00446 &\pm& 0.00354 \\ \nonumber  
  \text{Subtraction} : 0.00004 &\pm& 0.00040 \\ \nonumber
  \text{Importance} : -0.00400 &\pm& 0.00300  
\end{eqnarray*}
so we do indeed find that subtraction reduces the variance significantly.
 
\begin{figure}[htp]
  \begin{center}
    \subfigure[Gaussian \(m=10\)]{%
      \includegraphics[angle=-90,width=0.45\textwidth]{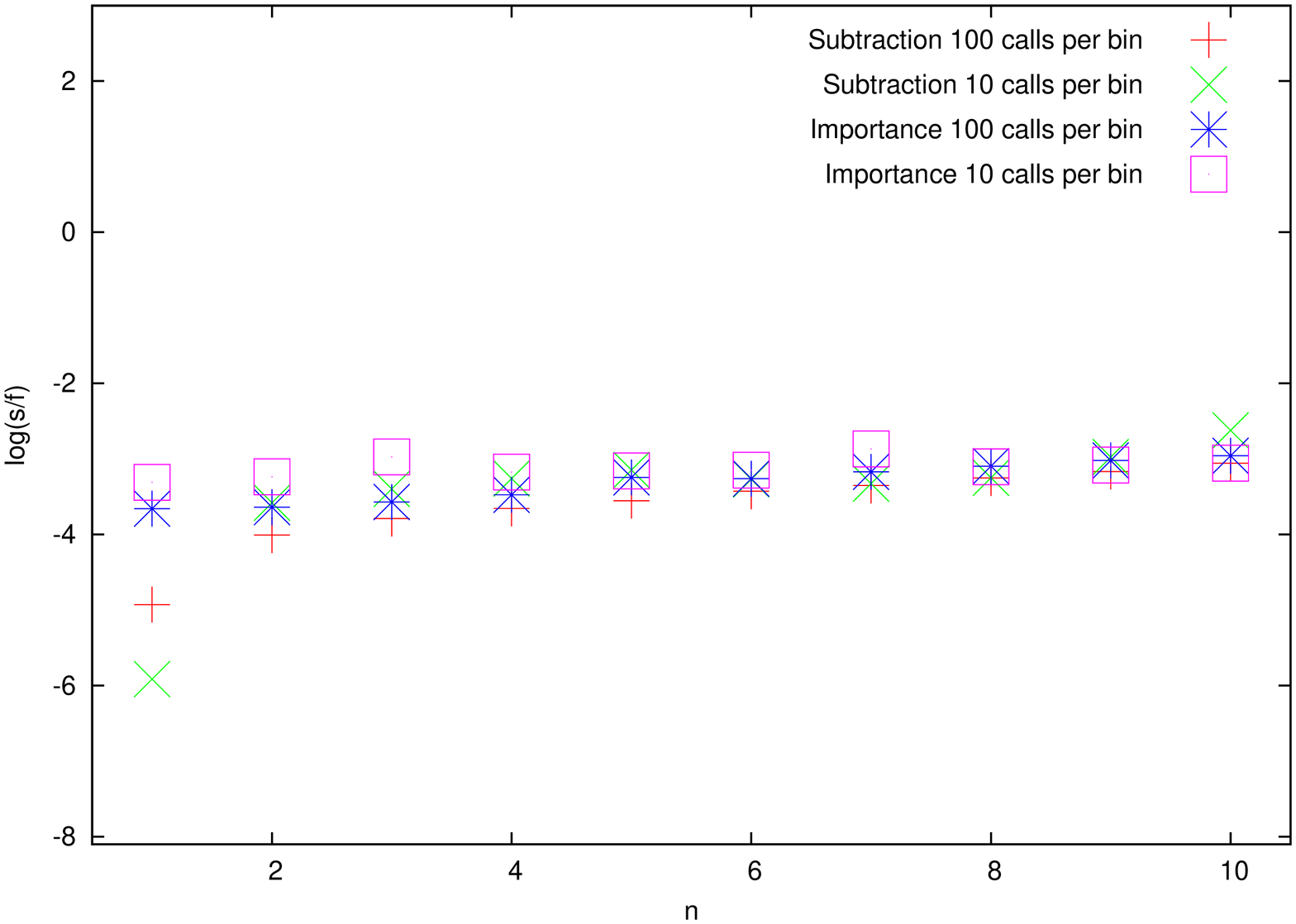}} 
    \subfigure[Gaussian \(m=100\)]{%
      \includegraphics[angle=-90,width=0.45\textwidth]{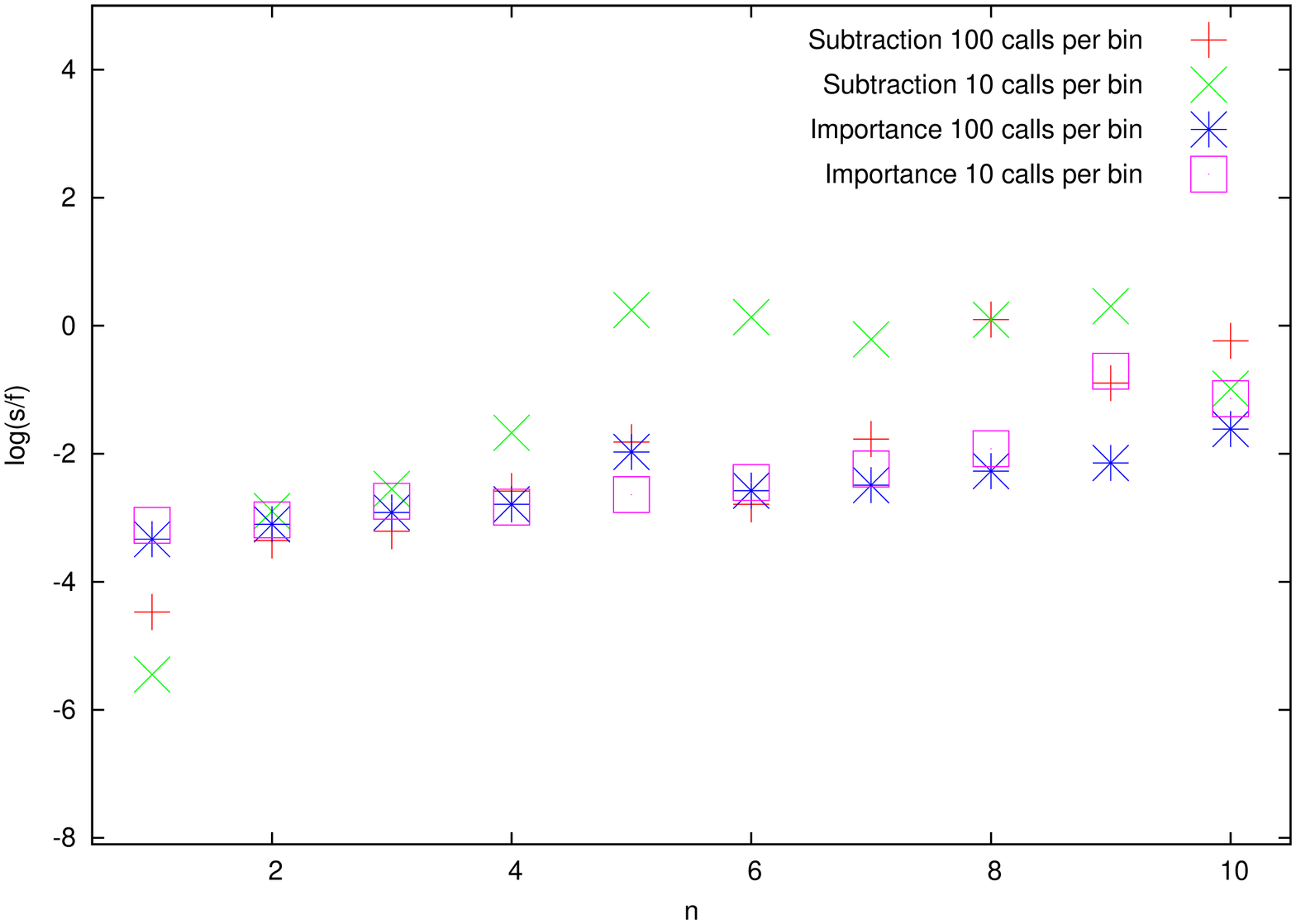}}
  \caption{Log of the standard error over the integral evaluation for equation
    \eqref{tst3} versus dimension using \(4\times10^4\) points on the grid
    obtained after four iterations of \(4\times10^4\) calls.  (a)~\(m=10\) the
    subtraction method works well in low dimensions.  (b)~\(m=100\) for this
    many function calls the subtraction method breaks down in dimension greater
    than \(5\) the separate integral estimates are inconsistent with large
    \(\chi^2\) and the error is large.  In contrast, importance sampling works
    well here.}
  \label{fig:exp1}  
  \end{center}
\end{figure}

\begin{figure}[htp]
  \begin{center}
    \subfigure[Gaussian \(m=10\)]{%
      \includegraphics[angle=-90,width=0.45\textwidth]{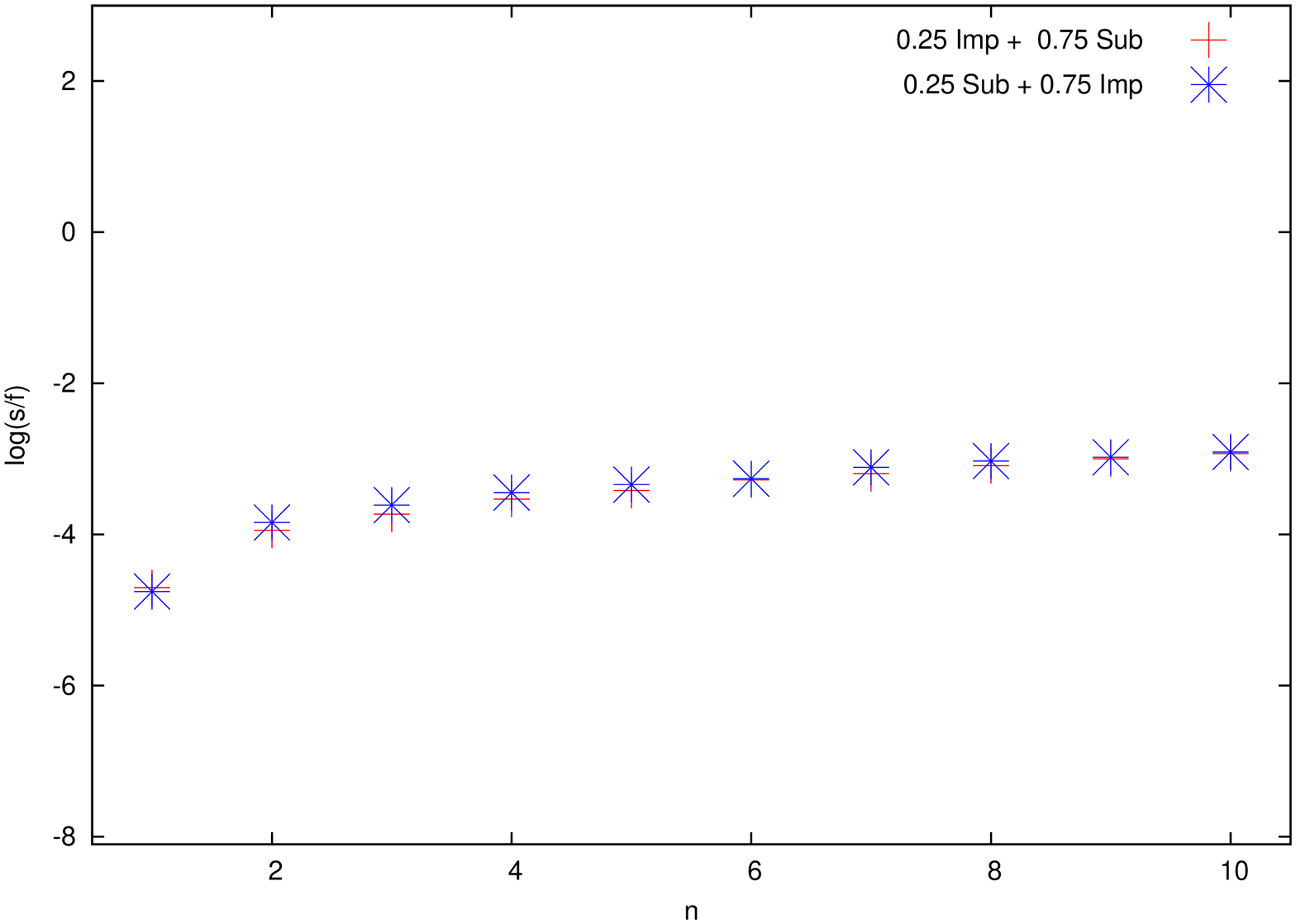}} 
    \subfigure[Gaussian \(m=100\)]{%
      \includegraphics[angle=-90,width=0.45\textwidth]{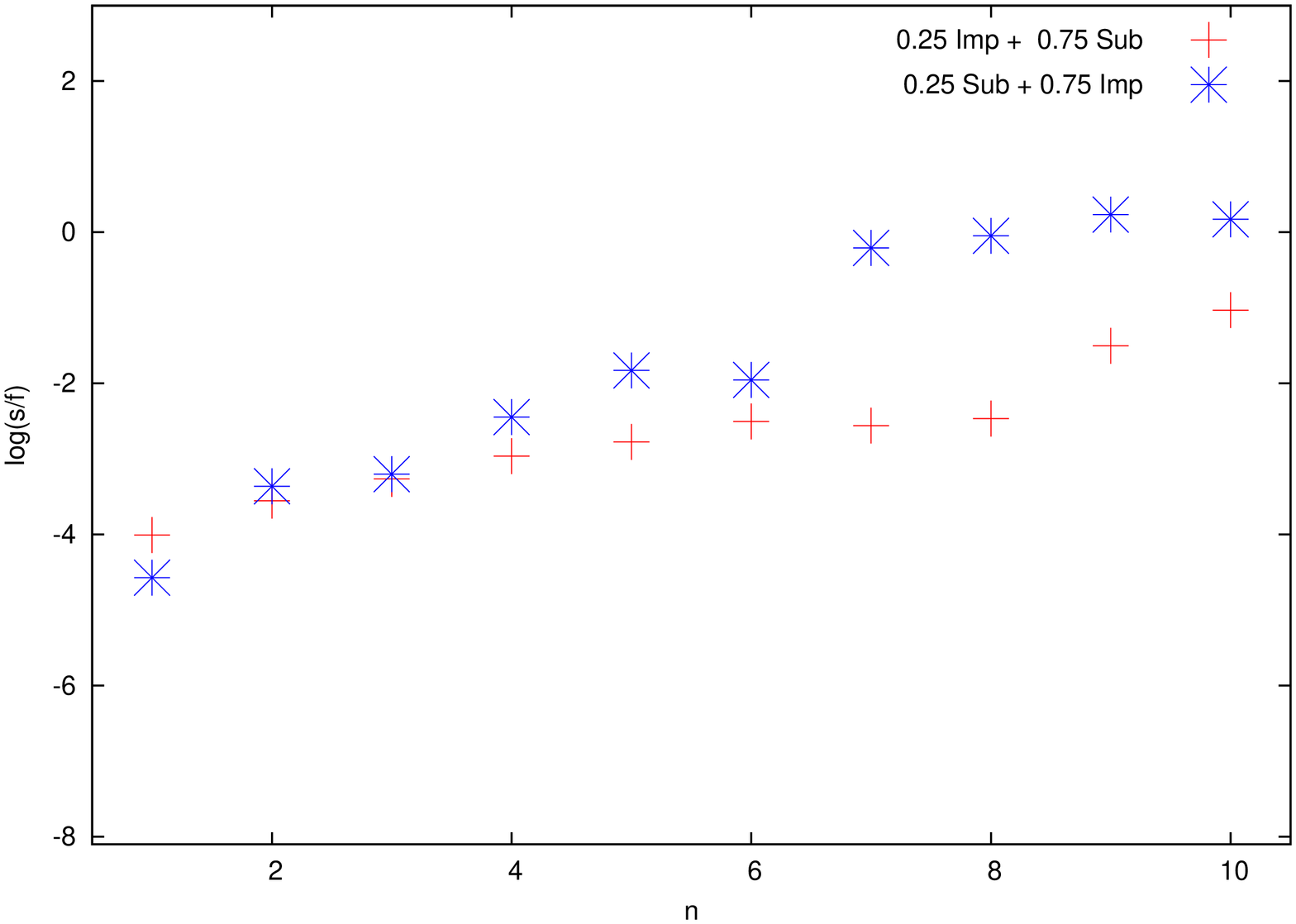}}
  \caption{Log of the standard error over the integral evaluation for equation
    \eqref{tst3} versus dimension using \(4\times10^4\) points on the grid
    obtained after four iterations of \(4\times10^4\) calls.  For both cases
    the mixed approach works well, retaining the advantage of subtraction in
    low dimensions and the stability of importance sampling in high.  Calls per
    bin = \(100\).}
  \label{fig:exp2} 
  \end{center}
\end{figure}

The integral
\begin{equation}
  \int_{[0,1]^d} dx_1\cdots dx_d \prod_i^n \sqrt{\frac m\pi}
    \exp\left(-m(x_i-\half) ^2\right) 
    = \left[\erf\left(\frac{\sqrt m}2\right)\right]^d
  \label{tst3}
\end{equation}
is another interesting case.  When the parameter \(m\) is small, the Gaussian
is broad and our integration scheme approximates the true value well.  However
as the dimension increases our method fails to find the peak of the Gaussian as
well as VEGAS and the integration starts to break down.  The subtraction method
can be improved by increasing the number of points per bin, however it still
does slightly worse than VEGAS.  This is possibly due to the interpolation
performed during the rebinning step failing to reconstruct the approximation
function.  There is no reason for subtraction to be better in this case and we
find, for a fixed number of calls as the dimension increases and therefore, the
number of bins per axis reduces for this function subtraction gets worse more
quickly than importance sampling.

Importance sampling will do better than subtraction in some cases and vice
versa therefore combining the two approaches may be beneficial.  We can do
subtraction for some percentage of our total number of function evaluations,
generating an approximation function and then do importance sampling on
\(f-\hat f\) for the remainder.  Or we could do an initial run of importance
sampling, then construct the approximation function on the bins with width
given by~\(\rho\).  Figure \ref{fig:exp2} shows the integral \eqref{tst3} in
various dimensions using both
varieties of mixed sampling.  The mixed sampling is done by using \(25\%\) of
the function evaluations on one kind of sampling and then using the grid and
approximation function found this way for the rest of the calls.  
The mixed strategies seem to
give good results reproducing the superiority of subtraction in low dimensions
and the stability of importance sampling in higher.

\begin{figure}[htb]
  \begin{center}
    \epsfxsize=100mm \epsffile{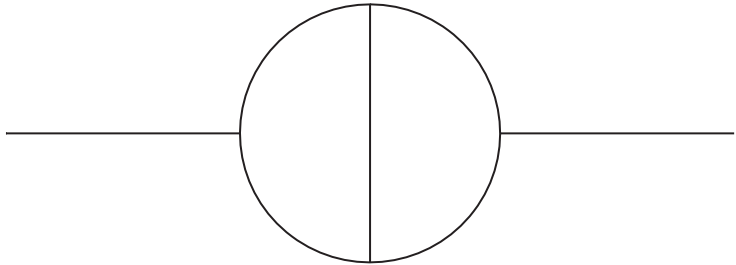}
  \end{center}
  \caption{\(\phi^3\) theory graph evaluated in the text.  Explicitly given in
    attached files.}
  \label{fig:phi}
\end{figure}

As an interesting practical example we perform the integral derived from the
Feynman parameterization of the scalar Feynman diagram shown in
Figure~\ref{fig:phi} in \(4\)-dimensional Euclidean space, and given explicitly
in a file attached to this paper.
\begin{figure}[htb]
  \begin{center}%
    \includegraphics[angle=0,width=0.75\textwidth]{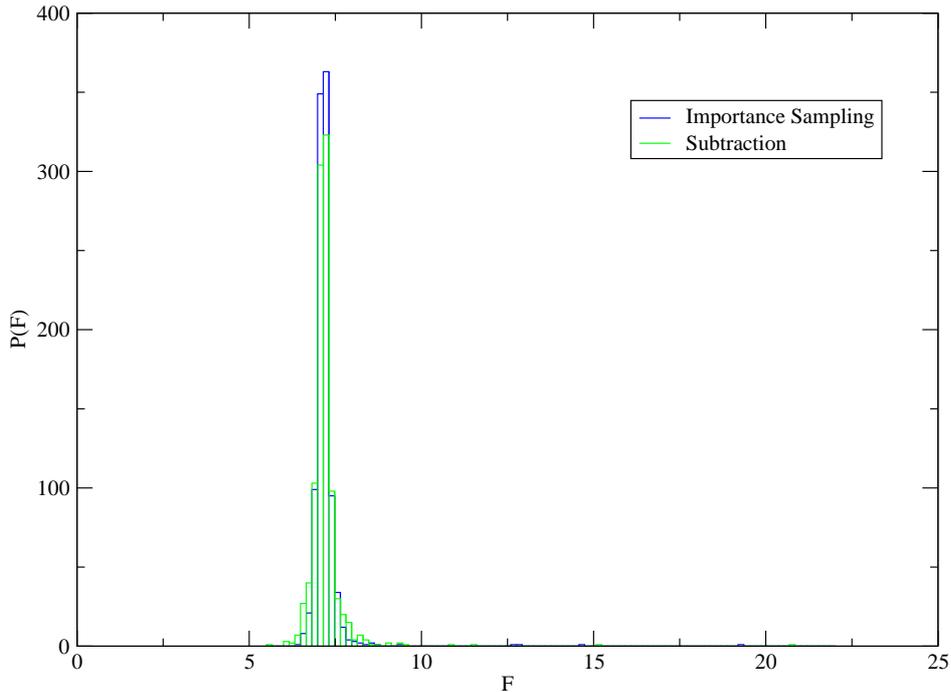}
  \end{center}
  \caption{ Distribution of results for the \(\phi^3\) graph.  The distribution
    is non-Gaussian in both cases with values differing significantly from the
    mean occuring far more often than if it were Gaussian --- the ``fat tail''
    of the distribution.}
  \label{fig:phi3}
\end{figure}
This integral is known to converge (its exact value is \(-7.2123414\ldots\))
but it is not necessarily square integrable and its higher moments may not
exist.  We plot the distribution obtained from evaluating the integral
\(1,000\) times by subtraction and importance sampling.  The distribution
looks distinctly non-Gaussian.  It is peaked around the exact value but were
the standard deviation used to estimate the error it would be incorrect.  This
kind of problem with parameter integrals arises often in practice and we
advise caution in interpreting the output of any Monte Carlo calculation.

\section{Conclusions}

We have reviewed Monte Carlo integration, pointing out some often overlooked
issues when the integral exists but the variance or some higher moments do
not.  This leads to non-Gaussian distributions and inaccurate estimates of the
error.  We have proposed a new algorithm for numerical Monte Carlo integration
based on adaptive subtraction.  We expect this method to have an advantage
over importance sampling Monte Carlo schemes for some integrands.  We have
implemented our proposed adaptive subtraction method and found it to be
better than importance sampling in some cases.  A
combination of the two methods can be better than either separately.  We have
given an explicit program, PANIC\footnote{Program for Adaptive Numerical
  Integration Computations} which is competitive with VEGAS.  PANIC also
shares the assumption of factorisability into a product of functions with
VEGAS, thus improvements to VEGAS that work around this assumption,
e.g.,~\cite{Ohl:1998jn,Hahn:2004fe}, will also work with subtraction and are
an interesting future direction.  The subtraction idea can work with any Monte
Carlo scheme and the subtraction function can even be constructed
independently, in this paper we have seen good results when combined with
importance sampling.  Our code, PANIC, is included.

\section*{Acknowledgements}

ADK would like to thank Benny Lautrup, who wrote one of the seminal papers in
the field \cite{Lautrup}, for useful discussions and for the acronym~PANIC.

\end{document}